\documentclass[10pt]{article}
\usepackage{graphicx}
\DeclareGraphicsExtensions{.pdf,.png,.jpg}
\usepackage{subfigure}
\usepackage{lineno}

\def\Title#1{\begin{center} {\Large #1 } \end{center}}
\def\Author#1{\begin{center}{ \sc #1} \end{center}}
\def\Address#1{\begin{center}{ \it #1} \end{center}}

\newcommand\pubblock{\rightline{\begin{tabular}{l} Proceedings of the Second Annual LHCP\\ \pubnumber\\
         \pubdate  \end{tabular}}}

\newenvironment{Abstract}{\begin{quotation} \begin{center} 
             \large ABSTRACT \end{center}\bigskip 
      \begin{center}\begin{large}}{\end{large}\end{center} \end{quotation}}

\newenvironment{Presented}{\begin{quotation} \begin{center} 
             PRESENTED AT\end{center}\bigskip 
      \begin{center}\begin{large}}{\end{large}\end{center} \end{quotation}}

\def\beq{\begin{equation}}
\def\eeq#1{\label{#1}\end{equation}}
\def\eeqn{\end{equation}}

\def\beqa{\begin{eqnarray}}
\def\eeqa#1{\label{#1}\end{eqnarray}}
\def\eeqan{\end{eqnarray}}

\let\bar=\overbar

\def\Dslash{\not{\hbox{\kern-4pt $D$}}}
\def\dslash{\not{\hbox{\kern-2pt $\del$}}}

\def\msb{{\bar{\ssstyle M \kern -1pt S}}}

\usepackage{color}

\newcommand\pubnumber{ ATL-PHYS-PROC-2014-122 }

\newcommand\pubdate{\today}

\def\affiliation{On behalf of the ATLAS Collaboration} 

\begin{document}
\large
\begin{titlepage}
\pubblock

\vfill
\Title{Searches for vector-like quarks and $t\bar{t}$ resonances with the ATLAS detector}
\vfill

\Author{B\'arbara \'Alvarez Gonz\'alez }
\Address{\affiliation}
\vfill
\begin{Abstract}
Searches for vector-like quarks and $t\bar{t}$ resonances are performed with the ATLAS experiment at the Large Hadron Collider using an integrated luminosity 
of 14.3~fb$^{-1}$ of proton-proton collisions recorded in 2012 at a center-of-mass energy of $\sqrt{s} = $ 8~TeV.
Several final states have been exploited to carry out these searches such as lepton plus jets and opposite and same-sign dilepton final states.
No significant excess of events above the Standard Model expectation is observed, and upper limits at 95$\%$ CL are derived for vector-like quarks of various
masses in a two-dimensional plane of branching ratios, and for $t\bar{t}$ resonances in two benchmark models, a topcolor leptophobic $Z'$ and a Kaluza-Klein gluon.
\end{Abstract}
\vfill

\begin{Presented}
The Second Annual Conference\\
 on Large Hadron Collider Physics \\
Columbia University, New York, U.S.A \\ 
June 2-7, 2014
\end{Presented}
\vfill
\end{titlepage}
\def\thefootnote{\fnsymbol{footnote}}
\setcounter{footnote}{0}
%

\normalsize 


\section{Introduction}
The ATLAS detector~\cite{detector} is a multi-purpose particle physics detector at the Large Hadron Collider (LHC)
at CERN. It is optimized to record information coming from proton-proton collisions happening in the center of the
detector. One of the primary goals of experiments at the LHC is the search for new physics beyond the Standard
Model (SM), which could solve some of the limitation of the SM like the nature of the symmetry breaking mechanism.
One alternative is vector-like quarks defined as quarks for which both chiralities have the same transformation 
properties under the electroweak gauge group. These heavy quarks are features of several models beyond the SM in which 
the electroweak symmetry is broken dynamically by a new strong interaction, such as in Topcolor~\cite{topcolor1, topcolor2}, 
Little Higgs~\cite{LittleH1, LittleH2}, and Composite Higgs~\cite{CompositeH} scenarios.
Some theories predict new heavy bosons that decay primarily into $t\bar{t}$ pairs. Examples of these theories include 
topcolor models, and Randall-Sundrum models with warped extra dimensions~\cite{KK}.

Searches for vector-like quarks and $t\bar{t}$ resonances are performed with the ATLAS experiment at the LHC 
using an integrated luminosity of 14.3~fb$^{-1}$ of proton-proton collisions recorded in 2012 at a center-of-mass 
energy of $\sqrt{s} = $ 8~TeV.  The search for pair production of a heavy top-like quark decaying to a 
high-$p{_\mathrm{T}}$ $W$~boson and a $b$ quark is presented in Section~\ref{Wboson}. Section~\ref{Zboson} describes the search for vector-like quarks that 
decay to a $Z$~boson and a third generation quark.
In Section~\ref{Higgs}, the vector-like quarks decaying to a Higgs boson and a top quark searches are presented.
Section~\ref{SameSign} describes searches containing same-sign dilepton events and, finally, Section~\ref{ttbar} presents the search for $t\bar{t}$ resonances.
No significant excess of events above the SM expectation is observed, and upper limits at 95$\%$ confidence level (CL) are derived for 
vector-like quarks of various masses in a two-dimensional plane of branching ratios and for $t\bar{t}$ resonances in two benchmark models, a topcolor leptophobic $Z'$ and a Kaluza-Klein gluon.

\section{\boldmath Search for pair production of a heavy top-like quark decaying to a high-$p{_\mathrm{T}}$ $W$~boson and a $b$~quark}
\label{Wboson}
A search is presented for production of a heavy top-like quark together with its
antiparticle ($T\bar{T}$), assuming a significant branching ratio decaying into a $W$~boson and
a $b$~quark~\cite{Wboson}. This analysis is performed in the lepton plus jets final state with a high-$p{_\mathrm{T}}$ isolated 
electron or muon, large missing transverse momentum, $E_\mathrm{T}^{miss}$, and more than three jets but less than six (to keep 
orthogonality to the search described in Section~\ref{Higgs}), with at least one of them  identified as originating 
from a $b$-quark ($b$-tagged).   
The key strategy of this analysis is to identify the hadronically-decaying $W$~boson, $W_{had}$, as well as the $b$-tagged 
jets ($b$-jets) in the event.
Two types of $W_{had}$ candidates are defined depending on the angular separation between their decay products.
The candidate $b$-jets are defined as the two jets with the highest $b$-tag discriminant.
The reconstructed mass of the $T$ quark is the final discriminating variable used in this search, it is built
from the $W_{had}$ candidate and one of the two $b$-jet candidates. The combination of the $b$-jet and the $W$ candidate 
is chosen to be the one with the smallest difference between the two reconstructed heavy quark masses.

The result of this search is interpreted both in the context of a chiral fourth-generation $T$ quark, as well as more
generically in the context of vector-like quark models.
For a chiral fourth generation quark, and under the assumption of
a branching ratio BR($T \rightarrow Wb$) = 1, a mass lower than 740~GeV is excluded at 95$\%$ CL.
For a vector-like singlet $T$ quark, the observed 95$\%$ CL limit is $m_T > $ 505~GeV.
For vector-like $T$ quarks, and under the assumption that only the $T \rightarrow Wb$, $T \rightarrow Zt$ and
$T \rightarrow Ht$ decay modes contribute, 95$\%$ CL upper limits are derived for various masses in
the two-dimensional plane of BR($T \rightarrow Wb$) versus BR($T \rightarrow Ht$), where $H$ is the SM Higgs boson. 
These limits are significantly improved by combining this search with the search described in Section~\ref{Higgs}.

\section{\boldmath Search for pair production of new heavy quarks that decay to a $Z$~boson and a third generation quark}
\label{Zboson}
\begin{figure}[h]
\centering
\includegraphics[width=0.45\textwidth, clip=]{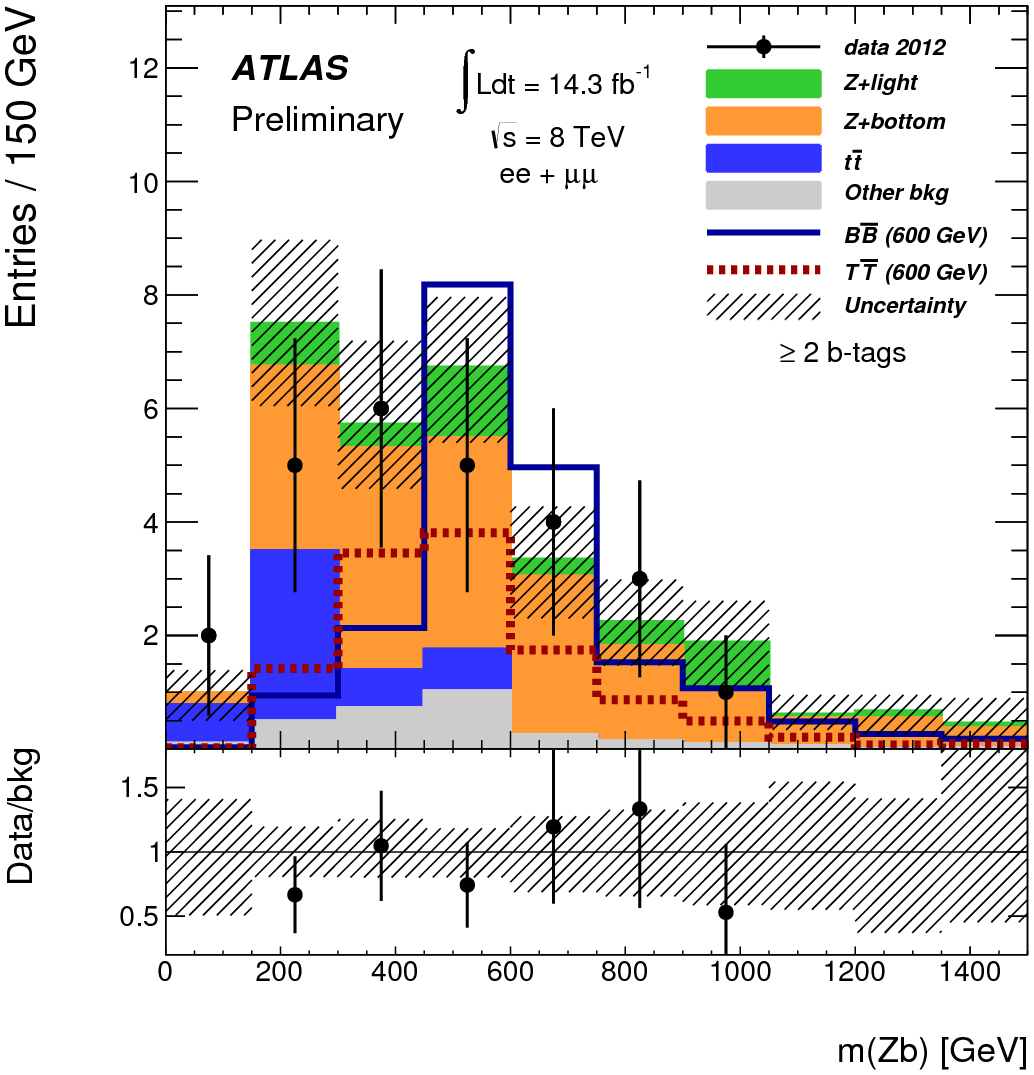}
\caption{The $m(Zb)$ distribution after final selection, in $Z$ candidate events containing at least two $b$-tagged jets. The hatched bands 
in the upper and lower panels represent the total background uncertainty. The rightmost bin in each histogram contains overflow events~\cite{Zboson}.}
\label{fig:Zboson}
\end{figure}

A search for the production of a new heavy quark with its antiparticle is presented, assuming
the new quark has a significant branching ratio to a $Z$~boson and a third generation SM quark~\cite{Zboson}, 
$T \rightarrow Zt$ and $B \rightarrow Zb$. 
Selected events contain a high transverse momentum $Z$~boson candidate reconstructed from a pair of oppositely
charged electrons or muons, at least two $b$-tagged jets and large total transverse momentum of all central jets.
After the final event selection (requiring $p{_\mathrm{T}}(Z) >$ 150~GeV and the scalar sum of the $p{_\mathrm{T}}$ of the selected 
jets, $H_\mathrm{T}$(jets), higher than 600~GeV), the invariant mass of 
the $Z$~boson candidate and the $b$-tagged jet with the highest transverse momentum, $m(Zb)$, is tested for a signal-like 
excess of events beyond the SM prediction, see Figure~\ref{fig:Zboson}.
Upper limits are derived for vector-like quarks of various masses in a two-dimensional plane of branching ratios. 
Under branching ratio assumptions corresponding to a weak-isospin singlet scenario, a T (B) quark with mass 
lower than 585 (645)~GeV is excluded at the 95$\%$ CL. Under branching ratio assumptions
corresponding to a particular weak-isospin doublet scenario, a T (B) quark with mass lower
than 680 (725)~GeV is excluded at the 95$\%$ CL.

\section{\boldmath Search for heavy top-like quarks decaying to a Higgs boson and a top quark}
\label{Higgs}

\begin{figure}[h]
\centering
\includegraphics[width=0.45\textwidth, clip=]{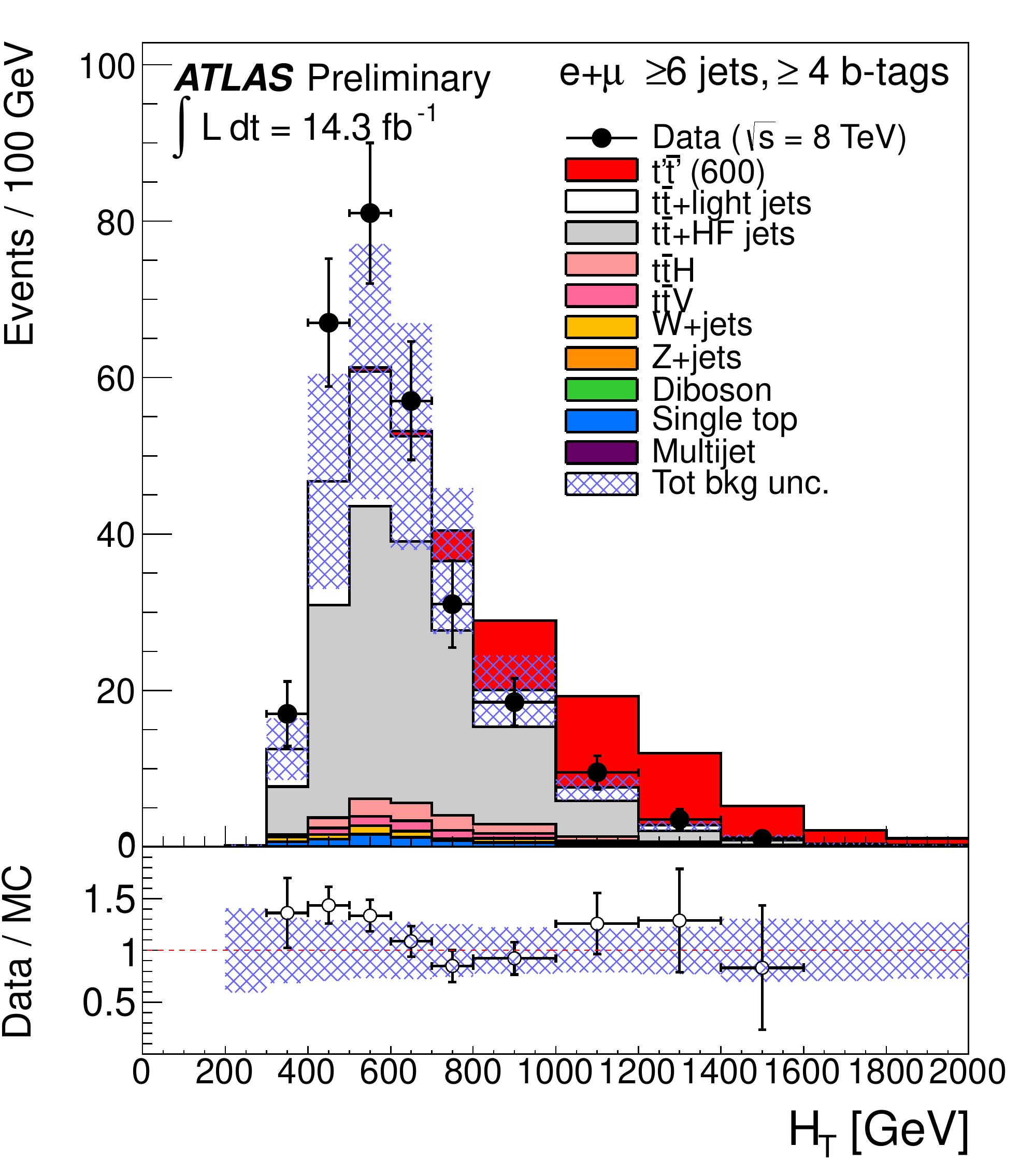}
\caption{Comparison between data and simulation for $H_\mathrm{T}$ in events with $\geq$ 6 jets and $\geq$ 4 $b$-tags. 
The $t\bar{t}$ background prediction is after fitting to data using the full $H_\mathrm{T}$ spectrum. Also shown is the expected $T \bar{T}$ signal corresponding to $m_T = $  600~GeV
in the $T$ doublet scenario. The bottom panel displays the ratio between data and background prediction. 
The shaded area represents the total post-fit background uncertainty~\cite{Higgs}.}
\label{fig:Ht}
\end{figure}

The production of a heavy up-type quark ($T$) together with its antiparticle, assuming a significant branching ratio 
decaying into a SM Higgs boson and a top quark as predicted by vector-like quark models, is also presented~\cite{Higgs}. 
Data are analyzed in the lepton plus jets final state,
characterized by an isolated electron or muon with moderately high transverse momentum,
significant missing transverse momentum, and at least six jets (orthogonal selection w.r.t the analysis described in Section~\ref{Wboson}). The search exploits the
high total transverse momenta of all final state objects and the high multiplicity of $b$-jets
characteristic of signal events with at least one SM Higgs boson decaying into $b\bar{b}$, to discriminate
against the dominant background from top quark pair production. Figure~\ref{fig:Ht} shows the $H_\mathrm{T}$ spectra 
used to derive 95$\%$ CL upper limits on the $T\bar{T}$  production cross section times branching fraction.

Under the branching ratio assumptions corresponding to a weakisospin doublet (singlet), an observed 95$\%$ CL limit $m_T >$~790~(640)~GeV is obtained. 
Upper limits are also derived for vector-like quarks of various masses in the two-dimensional plane of BR($T \rightarrow Wb$) versus
BR($T \rightarrow Ht$), where H is the SM Higgs boson, assumed to have a mass of 125~GeV. 
The branching ratio BR($T \rightarrow Zt$) is fixed to BR($T \rightarrow Zt$)~=~1 - BR($T \rightarrow Wb$) - BR($T \rightarrow Ht$). 
For instance, a $T$ quark with a mass of 600~GeV and BR($T \rightarrow Ht$)~$>$~0.3 is excluded at $\geq$ 95$\%$ CL,
regardless of the value of its branching ratios to $Wb$ and $Zt$.

\section{\boldmath Search for anomalous production of events with same-sign dileptons and $b$-jets}
\label{SameSign}
A search for exotic processes that result in final states containing jets (including at least one $b$-jet), 
sizable $E_\mathrm{T}^{miss}$, and a pair of leptons with the same electric charge is presented~\cite{SameSign}. 
This is a promising search channel since the SM yields of such events are small, and several types of new physics may contribute.
There are several models that predict an enhanced rate of production of such events beyond the expectations of the SM; 
the ones considered in this letter are pair production of chiral $b'$ quarks, pair production of vector-like quarks ($B\bar{B}$), 
enhanced four top quark production, same-sign top quark pair production, sgluon pair production (where sgluons are color adjoint 
scalars that appear in several extensions to the SM), and pair production of Kaluza-Klein excitations of the photon.  
A common dataset is used to search for each of these signatures, and the event selection is optimized for each signal model,
more details in Ref.~\cite{SameSign}. The 95$\%$ CL observed limits are summarized in Table~\ref{tab:table_samesing}.

  \begin{table}[h]
    \begin{center}
      \begin{tabular}{l|ccccccc}  
        \hline \hline
        Variable &  $m_{b'}$      &  $m_{B}$      & $m_{T}$    &$\sigma_{4-tops}$  & $\sigma_{SS-top}$ & $m_{sgluon}$  & $m_{KK}$\\ \hline
        Limits & $>$ 0.72~TeV  & $>$ 0.59~TeV & $>$ 0.54~TeV & $<$ 85 fb       &  $<$ 210 fb    & $>$ 0.80~TeV & $>$ 0.90~TeV\\
        \hline \hline
      \end{tabular}
      \caption{ 95$\%$ CL observed limits for the chiral $b'$ quark mass ($m_{b'}$), the vector-like quark $B$ mass ($m_{B}$), the vector-like quark $T$ mass ($m_{T}$), the four top quark production cross section ($\sigma_{4-tops}$), the same-sign top quark pair production cross section ($\sigma_{SS-top}$), the sgluon mass ($m_{sgluon}$)  and the Kaluza-Klein mass ($m_{KK}$). The limits on the mass of the $b'$ assumme 100$\%$ branching fraction to $Wt$.  
        For the vector-like quarks, the branching ratios to $W$, $Z$, and $H$ decay modes are assumed to be consistent with the $B$ or $T$ being a singlet.}
      \label{tab:table_samesing}
    \end{center}
  \end{table}

\section{\boldmath Search for $t\bar{t}$ resonances}
\label{ttbar}

The search for new particles that decay into top quark pairs ($t\bar{t}$) is carried out in the lepton plus 
jets final state ($t\bar{t} \rightarrow W^+bW^-\bar{b}$), where one $W$~boson decays leptonically and the other
 hadronically~\cite{ttbar}. The $t\bar{t}$ system is reconstructed using both a conventional resolved jet 
analysis and a large-radius jet substructure analysis. 
Using the former, the hadronically decaying top quark is identified by two or three distinct small-radius jets. 
Using the latter, the hadronically decaying top quark is identified by one large-radius jet that has substructure 
consistent with being composed of the decay products of a $W$~boson and a $b$~quark. 
High momentum top quark decays are reconstructed more efficiently using the boosted reconstruction technique. 
For both reconstruction methods, the semileptonically decaying top quark is identified by a lepton, one small-radius jet and $E_\mathrm{T}^{miss}$.

\begin{figure}[h]
\centering
\subfigure[]{\includegraphics[width=0.48\textwidth, clip=]{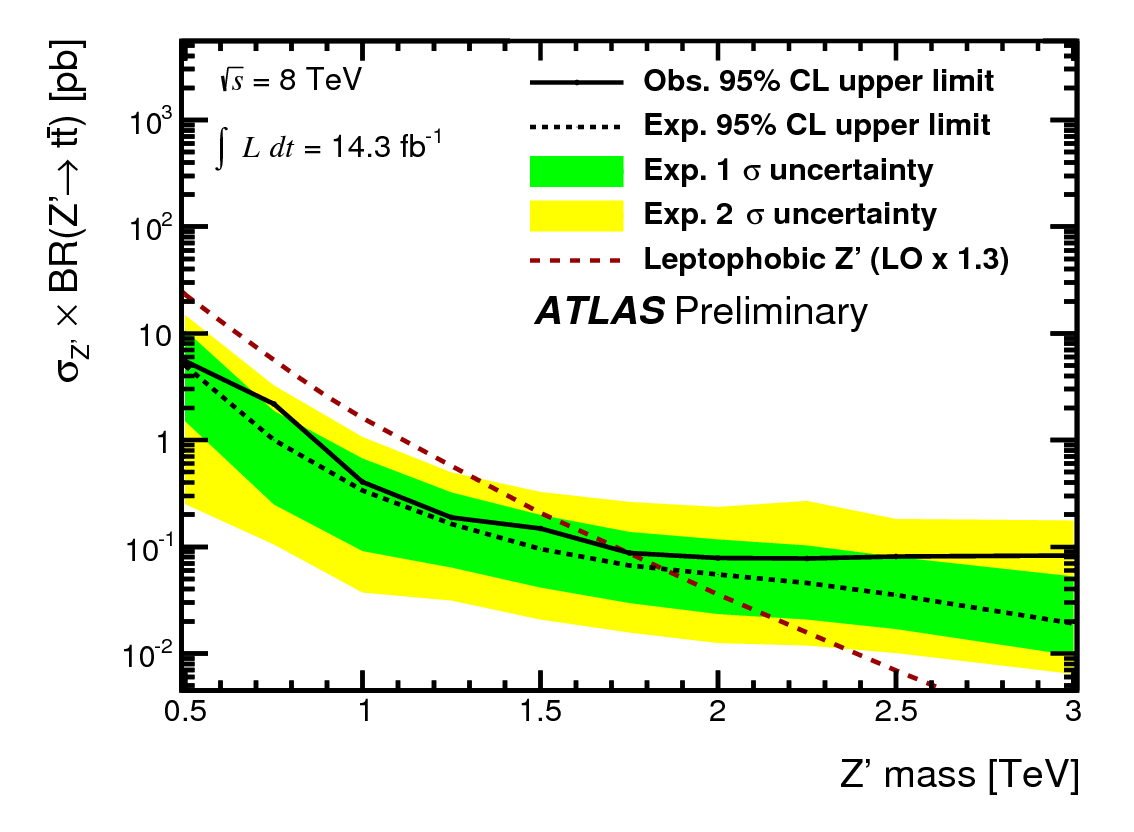}}
\subfigure[]{\includegraphics[width=0.48\textwidth, clip=]{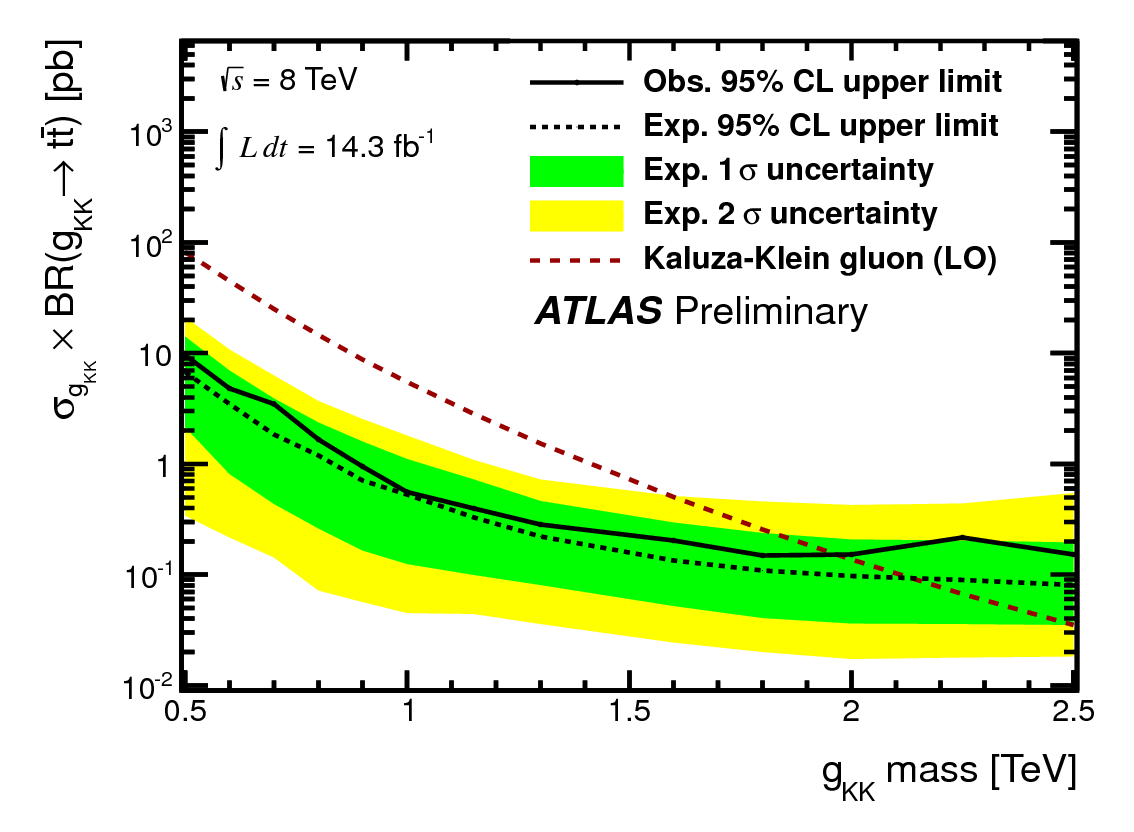}}
\caption{Observed and expected upper cross section limits times the $t\bar{t}$ branching ratio on (a) $Z'$
bosons and (b) Kaluza-Klein gluons. The resolved and the boosted selections have been combined
in the estimation of the limits. Both systematic and statistical uncertainties are included~\cite{ttbar}.}
\label{fig:figure_ttbar}
\end{figure}

The $t\bar{t}$ invariant mass spectrum is searched for local excesses deviating from the SM prediction. 
Many theories of new physics predict the existence of new particles, and some predict new heavy bosons that decay primarily 
into $t\bar{t}$ pairs. 
Two specific theoretical models as benchmarks are used in this search. The models test the production of resonances with both narrow and broad widths 
relative to the detector resolution which is of order 7$\%$: narrow resonance benchmark is a topcolor leptophobic $Z'$~\cite{Zprime} and broad resonance 
benchmarks are Kaluza-Klein (KK) gluons that arise in Randall-Sundrum models with an extra dimension with a warped geometry~\cite{KK}.
The 95$\%$ CL upper limits on the production rate are determined for massive states for these two benchmark models (see Figure~\ref{fig:figure_ttbar}) 
combining the four statistically uncorrelated spectra, corresponding to boosted and resolved selections, as well as e+jets and muon+jets decay channels.
This results in mass exclusion limits of a topcolor leptophobic $Z'$ boson between 0.5 and 1.8~TeV and a KK gluon between 0.5 and 2.0~TeV at 95$\%$ CL.

\section{Summary}
Searches in the lepton plus jets and opposite and same-sign dilepton decay channels have been carried out with the ATLAS experiment at the LHC. 
These searches use an integrated luminosity of 14.3~fb$^{-1}$ of proton-proton collisions recorded in 2012 at a center-of-mass energy of $\sqrt{s} = $ 8~TeV.
No significant excess of events above the Standard Model expectation is observed, and upper limits at 95$\%$ CL are derived for vector-like quarks of various
masses in a two-dimensional plane of branching ratios, and for $t\bar{t}$ resonances in two benchmark models, a topcolor leptophobic $Z'$ and a Kaluza-Klein gluon.

\end{document}